# Urban Spatial Order: Street Network Orientation, Configuration, and Entropy

Geoff Boeing

Department of Urban Planning and Spatial Analysis
Sol Price School of Public Policy
University of Southern California
boeing@usc.edu

August 2019

## Abstract

Street networks may be planned according to clear organizing principles or they may evolve organically through accretion, but their configurations and orientations help define a city's spatial logic and order. Measures of entropy reveal a city's streets' order and disorder. Past studies have explored individual cases of orientation and entropy, but little is known about broader patterns and trends worldwide. This study examines street network orientation, configuration, and entropy in 100 cities around the world using OpenStreetMap data and OSMnx. It measures the entropy of street bearings in weighted and unweighted network models, along with each city's typical street segment length, average circuity, average node degree, and the network's proportions of four-way intersections and dead-ends. It also develops a new indicator of orientation-order that quantifies how a city's street network follows the geometric ordering logic of a single grid. A cluster analysis is performed to explore similarities and differences among these study sites in multiple dimensions. Significant statistical relationships exist between city orientation-order and other indicators of spatial order, including street circuity and measures of connectedness. On average, US/Canadian study sites are far more grid-like than those elsewhere, exhibiting less entropy and circuity. These indicators, taken in concert, help reveal the extent and nuance of the grid. These methods demonstrate automatic, scalable, reproducible tools to empirically measure and visualize city spatial order, illustrating complex urban transportation system patterns and configurations around the world.

# Introduction

Spatial networks such as streets, paths, and transit lines organize the human dynamics of complex urban systems. They shape travel behavior, location decisions, and the texture of the urban fabric (Jacobs, 1995; Levinson and El-Geneidy, 2009; Parthasarathi et al., 2015). Accordingly, researchers have recently devoted much attention to street network patterns, performance, complexity, and configuration (Barthelemy et al., 2013; Batty, 2005a; Boeing, 2018a; Buhl et al., 2006; Chan et al., 2011; Ducruet and Beauguitte, 2014; Jiang et al., 2014; Jiang and Claramunt, 2004; Marshall, 2004; Masucci et al., 2013; Nilsson and Gil, 2019; Tsiotas and Polyzos, 2018; Wang, 2015). In these spatial networks, entropy has deep theoretical connections with complexity (Batty, 2005b; Batty et al., 2014). One research stream has explored the nature of entropy and order in urban street networks, seeking to quantify patterns of spatial order and disorder in urban circulation systems (Gudmundsson and Mohajeri, 2013; Li et al., 2018; Mohajeri et al., 2013a, 2013b; Mohajeri and Gudmundsson, 2014, 2012; Yeh and Li, 2001).

     Theories of urban order span sociological frameworks of physical-social disorder (e.g., "broken windows" theory), to public health goals of opening-up and sanitizing pathogenic urban spaces, to city planners' pursuit of functional differentiation and regulation (Boyer, 1983; Hatuka and Forsyth, 2005; Mele, 2017; O'Brien et al., 2019; Park and Burgess, 1925; Xu, 2008). This study considers the spatial logic and geometric ordering that arise through street network orientation. A city's development eras, design paradigms, underlying terrain, culture, and local economic conditions influence the pattern, topology, and grain of its street networks (Jackson, 1985; Kostof, 1991). These networks in turn structure the human interactions and transportation processes that run along them, forming an important pillar of city planners' quest for spatial order (Rose-Redwood and Bigon, 2018). In particular, network orientation and geometry have played an outsized role in urban planning since its earliest days (Smith, 2007).

     Measuring these network patterns can help researchers, planners, and community members understand local histories of urban design, transportation planning, and morphology; evaluate existing transportation system patterns and configurations; and explore new infrastructure proposals and alternatives. It also furthers the science of cities by providing a better understanding of urban patterns and how they correspond to evolutionary mechanisms, planning, and design. However, due to traditional data gathering challenges, this research literature has necessarily relied on small samples, limited geographies, and abstract indicators. Past studies have typically explored circuity and entropy in individual or paired case studies—less is known about broader cross-sectional trends worldwide. How do street network configurations organize and order urban space in cities around the world?

     This paper addresses this gap by empirically modeling and measuring order and configuration in 100 city street networks around the world, comprising over 4.8 million nodes and 3.3 million edges. It measures street network orientation entropy, circuity, connectedness, and grain. It also develops a straightforward new indicator, the orientation-order $\varphi$, to quantify the extent to which a street network follows the spatial ordering logic of a single grid. It finds significant statistical relationships between city orientation and other indicators of spatial order (including street circuity and connectedness). The most common orientation worldwide, even among cities lacking a strong grid, tends toward north-south-east-west. It also finds that American cities tend to be far more grid-like and less circuitous than cities

elsewhere. Considered jointly, this collection of indicators helps reveal the extent and nuance of the grid around the world.

# Background

## Street Network Planning

The orthogonal grid, the most common planned street pattern, is often traced back to Hippodamus of Miletus (Mazza, 2009; Paden, 2001)—whom Aristotle labeled the father of city planning for his orthogonal design of Piraeus in ancient Greece—but archaeologists have found vestiges in earlier settlements around the world (Burns, 1976; Stanislawski, 1946). Mohenjo-Daro in the Indus Valley, dating to 2500 BCE, featured a north-south-east-west orthogonal grid (McIntosh, 2007). Ancient Chinese urban design organized capital cities around gridded patterns codified in the Kao Gong Ji, a scientific text from c. 500 BCE (Elman and Kern, 2009). Teotihuacan featured an offset grid, dating to 100 BCE, that aligned with the Valley of Mexico's zenith sunrise (Peterson and Chiu, 1987; Sparavigna, 2017). The Roman Empire used standardized street grids to efficiently lay out new towns and colonies during rapid imperial expansion (Kaiser, 2011). Many medieval towns were even planned around approximate, if distorted, grids possibly to maximize sun exposure on east-west streets during winter market days (Lilley, 2001). In 1573, King Phillip II of Spain issued the Law of the Indies, systematizing how colonists sited new settlements and designed rectilinear gridded street networks around central plazas (Low, 2009; Rodriguez, 2005). In the US, many east coast cities planned their expansions around street grids, including Philadelphia in 1682, Savannah in 1733, Washington in 1791, and New York in 1811 (Jackson, 1985; Sennett, 1990). The subsequent US Homestead Act sweepingly organized the American interior according to the spatial logic of the gridiron (Boeing, 2018b).

 In the context of urban form, the concept of "spatial order" is fuzzy. Street networks that deviate from griddedness inherently possess different spatial logics and ordering principles (Karimi, 1997; Southworth and Ben-Joseph, 2004, 1995). Cities planned without a grid—as well as unplanned cities that grew through accretion—may lack clearly defined orientation order, but can still be well-structured in terms of complex human dynamics and land use (Hanson, 1989). Specific visual/geometric order should not be confused for functional/social order (Roy, 2005; Salingaros, 1998; Smith, 2007). Different design logics support different transportation technologies and appeal to different cultures and eras (Jackson, 1985).

 The grid has been used to express political power, promote military rule, improve cadastral legibility, foster egalitarianism, and encourage land speculation and development (Ellickson, 2013; Groth, 1981; Low, 2009; Martin, 2000; Mazza, 2009; Rose-Redwood, 2011; Sennett, 1990). Many cities spatially juxtapose planned and unplanned districts or non-binarily intermingle top-down design with bottom-up self-organized complexity. Old cores may comprise organic patterns adjacent to later gridirons, in turn adjacent to later winding suburbs. Even previously highly-ordered urban cores can grow in entropy as later generations carve shortcuts through blocks, reorganize space through infill or consolidation, and adapt to shifting points of interest—all of which occurred in medieval Rome and Barcelona, for instance (Kostof, 1991).

## Street Network Modeling

Street networks are typically modeled as graphs where nodes represent intersections and dead-ends, and edges represent the street segments that link them (Barthelemy and Flammini, 2008; Cardillo et al., 2006; Lin and Ban, 2013; Marshall et al., 2018; Porta et al., 2006). These edges are spatially embedded and have both a length and a compass bearing (Barthelemy, 2011). The present study models urban street networks as undirected nonplanar multigraphs with possible self-loops. While directed graphs most-faithfully represent constraints on flows (such as vehicular traffic on a one-way street), undirected graphs better model urban form by corresponding 1:1 with street segments (i.e., the linear sides of city blocks). While many street networks are approximately planar (having relatively few overpasses or underpasses), nonplanar graphs provide more accurate models by accommodating those bridges and tunnels that do often exist (Boeing, 2018c; Eppstein and Goodrich, 2008).

The data to study these networks typically come from shapefiles of digitized streets. In the US, the Census Bureau provides TIGER/Line shapefiles of roads nationwide. In other countries, individual municipal, state, or federal agencies may provide similar data, however, digitization standards and data availability vary. Accordingly, cross-sectional research of street network orientation and entropy has tended to be limited to individual geographical regions or examine small samples. However, today, OpenStreetMap presents a new alternative data source. OpenStreetMap is a collaborative worldwide mapping project that includes streets, buildings, amenities, and other spatial features. Although its data quality varies somewhat between countries, in general its streets data are high quality, especially in cities (Barrington-Leigh and Millard-Ball, 2017; Barron et al., 2014; Zielstra et al., 2013). This data source offers the opportunity to conduct cross-sectional research into street network form and configuration around the world.

Recently, scholars have studied street network order and disorder through circuity and orientation entropy. The former measures street curvature and how this relates to other urban patterns and processes (Boeing, 2019; Giacomin and Levinson, 2015; Levinson and El-Geneidy, 2009). The latter quantifies and visualizes the entropy of street orientations to assess how ordered they are (Courtat et al., 2011; Gudmundsson and Mohajeri, 2013; Mohajeri et al., 2013a, 2013b; Mohajeri and Gudmundsson, 2014, 2012), as entropy quantifies the fundamentally related concepts of disorder, uncertainty, and dispersion. Louf and Barthelemy (2014) explore city block geometries around the world as a function of block size and form factor, clustering them to identify differences between US and European cities. However, less is known about cross-sectional trends in the spatial orientation and ordering of street networks worldwide. This study builds on this prior research into circuity, order, and entropy by drawing on OpenStreetMap data to examine cities around the world and explore their patterns and relationships.

# Methods

## Data

To better understand urban spatial order and city street network entropy, we analyze 100 large cities across North America, South America, Europe, Africa, Asia, and Oceania. Our sampling strategy emulates Louf and Barthelemy's (2014) to select cities through a balance of high population, regional significance, and some stratification to ensure geographical diversity within regions. Accordingly, this sample comprises a broad cross-section of different histories, cultures, development eras, and design paradigms. Of course, no single consistent definition of "city" or its spatial jurisdiction exists worldwide as these vary between countries for historical and political reasons. We aim for consistency by trying to use each study site's closest approximation of a "municipality" for the city limits. The lone exception is Manhattan, where we focus on one borough's famous grid instead of the amalgam of boroughs that compose New York City.

Once these study sites are defined, we use the OSMnx software to download the street network within each city boundary and then calculate several indicators. OSMnx is a free, open-source, Python-based toolkit to automatically download spatial data (including municipal boundaries and streets) from OpenStreetMap and construct graph-theoretic objects for network analysis (Boeing, 2017).

## Analysis

For each city, we calculate the street network's edges' individual compass bearings with OSMnx using two different methods. The first method simplifies the topology of each graph such that nodes exist only at intersections and dead-ends; edges thus represent street segments (possibly curving, as full spatial geometry is retained) between them (*ibid.*). In this method, the bearing of edge $e_{uv}$ equals the compass heading from $u$ to $v$ and its reciprocal (e.g., if the bearing from $u$ to $v$ is 90° then we additionally add a bearing of 270° since the one-dimensional street centerline points in both directions). This captures the orientation of street segments but ignores the nuances of mid-block curvature. To address this, the second method does not simplify the topology: edges represent OpenStreetMap's raw straight-line street segments, either between intersections or in chunks approximating curving streets. This method weights each edge's bearing by length to adjust for extremely short edges in these curve-approximations. In both methods, self-looping edges have undefined bearings, which are ignored.

Once we have calculated all of the bearings (and their reciprocals) for all the edges in a city, we divide them into 36 equal-sized bins (i.e., each bin represents 10°). To avoid extreme bin-edge effects around common values like 0° and 90°, we shift each bin by -5° so that these values sit at the centers of their bins rather than at their edges. This allows similar common bearings such as 359.9° and 0.1° to fall in the same bin as each other. Once the bearings are binned, we calculate the Shannon entropy, $H$, of the city's orientations' distribution (Shannon, 1948). For each city's graph, we first calculate the entropy of the unweighted/simplified street orientations, $H_o$, as:

$$H_o = -\sum_{i=1}^{n} \text{P}(o_i) \log_e \text{P}(o_i) \tag{1}$$

where *n* represents the total number of bins, *i* indexes the bins, and P($o_i$) represents the proportion of orientations that fall in the *i*th bin. We similarly calculate the entropy of the weighted/unsimplified street orientations, $H_w$, as:

$$H_w = -\sum_{i=1}^{n} \text{P}(w_i) \log_e \text{P}(w_i) \tag{2}$$

where *n* represents the total number of bins, *i* indexes the bins, and P($w_i$) represents the proportion of weighted orientations that fall in the *i*th bin. While $H_w$ is biased by the city's shape (due to length-weighting), $H_o$ is not.

The natural logarithm means the value of $H$ is in dimensionless units called "nats," or the natural unit of information. The maximum entropy, $H_{max}$, that any city could have equals the logarithm of the number of bins: 3.584 nats. This represents the maximum entropy distribution, a perfectly uniform distribution of street bearings across all bins. If all the bearings fell into a single bin, entropy would be minimized and equal 0. However, given the undirected graph, the minimal theoretical entropy a street network could have (e.g., if all of its streets ran only north-south, thus falling evenly into two bins) would be 0.693 nats. But given the nature of the real world, a more plausible minimum would instead be an idealized city grid with all streets in four equal proportions (e.g., north-south-east-west). This perfect grid entropy, $H_g$, would equal 1.386 nats. Therefore, we can calculate a normalized measure of orientation-order, $\varphi$, to indicate where a city stands on a linear spectrum from completely disordered/uniform to perfectly ordered/grid-like as:

$$\varphi = 1 - \left( \frac{H_o - H_g}{H_{max} - H_g} \right)^2 \tag{3}$$

Thus, a $\varphi$ value of 0 indicates low order (i.e., perfect disorder and maximum entropy with a uniform distribution of streets in every direction) and a $\varphi$ value of 1 indicates high order (i.e., a single perfectly-ordered idealized four-way grid and minimal possible entropy). Note that the value is squared to linearize its already normalized scale between 0 and 1, allowing us to interpret it as the *extent* to which a city is ordered according to a single grid. All remaining indicators' formulae use the (unweighted) simplified graph for the most faithful model of the urban form, geographically and topologically. We calculate each city's median street segment length $\tilde{l}$, average node degree $\bar{k}$ (i.e., how many edges are incident to the nodes on average), proportion of nodes that are dead-ends $P_{de}$, and proportion of nodes that are four-way intersections $P_{4w}$. Finally, we calculate each city street network's average circuity, $\varsigma$, as:

$$\varsigma = \frac{L_{net}}{L_{gc}} \tag{4}$$

where $L_{net}$ represents the sum of all edge lengths in the graph and $L_{gc}$ represents the sum of all great-circle distances between all pairs of adjacent nodes. Thus, $\varsigma$ represents how much more circuitous a city's street network is than it would be if all its edges were straight-line paths between nodes (Boeing, 2019; Qureshi et al., 2002).

We visualize these characteristics and examine their statistical relationships to explore the nature of spatial order/disorder in the street networks' orientations, hypothesizing that more-gridded cities (i.e., higher $\varphi$ values) have higher connectedness (i.e., higher node degrees, more four-way intersections, fewer dead-ends) and less-winding street patterns. Finally, to systematically interpret city similarities and differences, we cluster the study sites in a four-dimensional feature space of the key indicators of interest ($\bar{k}$, $\varphi$, $\tilde{l}$, and $\varsigma$), representing a cross-section of street network character. We first standardize the features for appropriate scaling, then perform hierarchical agglomerative clustering using the Ward linkage method with a Euclidean metric.

## Results

Table 1 presents the indicators' values for each of the cities studied. We find that $H_o$ and $H_w$ are very strongly correlated (Pearson product-moment correlation coefficient $r > 0.99$, $p < 0.001$) and thus provide essentially redundant statistical information about these networks. Therefore, the remainder of these findings focus on $H_o$ unless otherwise explicitly stated. Three American cities (Chicago, Miami, and Minneapolis) have the lowest orientation entropies of all the cities studied, indicating that their street networks are the most ordered. In fact, all 16 cities with the lowest entropies are in the US and Canada. Outside of the US/Canada, Mogadishu, Kyoto, and Melbourne have the lowest orientation entropies. Surprisingly, the city with the highest entropy, Charlotte, is also in the US. São Paulo and Rome immediately follow it as the next highest cities. Chicago, the most ordered city, has a $\varphi$ of 0.90, while Charlotte, the most disordered, has a $\varphi$ of 0.002. Recall that a $\varphi$ of 0 indicates a uniform distribution of streets in every direction and a $\varphi$ of 1 indicates a single perfectly-ordered grid. Charlotte's and São Paulo's street orientations are nearly perfectly disordered.

**Table 1.** Resulting indicators for the 100 study sites.

| Region | City | $\varphi$ | $H_o$ | $H_w$ | $\tilde{l}$ | $\varsigma$ | $\bar{k}$ | $P_{de}$ | $P_{4w}$ |
|---|---|---|---|---|---|---|---|---|---|
| Asia/Oceania | Bangkok | 0.105 | 3.465 | 3.452 | 64.6 | 1.059 | 2.385 | 0.360 | 0.108 |
| | Beijing | 0.335 | 3.177 | 3.206 | 177.5 | 1.053 | 2.985 | 0.135 | 0.241 |
| | Hanoi | 0.010 | 3.573 | 3.572 | 64.4 | 1.065 | 2.610 | 0.246 | 0.102 |
| | Hong Kong | 0.012 | 3.571 | 3.563 | 61.0 | 1.137 | 2.932 | 0.114 | 0.174 |
| | Jakarta | 0.167 | 3.391 | 3.347 | 52.8 | 1.065 | 2.741 | 0.175 | 0.096 |
| | Kabul | 0.076 | 3.499 | 3.510 | 79.3 | 1.062 | 2.673 | 0.226 | 0.130 |
| | Karachi | 0.088 | 3.485 | 3.493 | 71.3 | 1.032 | 3.027 | 0.095 | 0.216 |
| | Kathmandu | 0.054 | 3.523 | 3.500 | 63.3 | 1.071 | 2.595 | 0.234 | 0.089 |
| | Kyoto | 0.357 | 3.148 | 3.229 | 49.6 | 1.090 | 2.887 | 0.134 | 0.157 |
| | Manila | 0.062 | 3.514 | 3.484 | 63.5 | 1.023 | 3.141 | 0.095 | 0.347 |
| | Melbourne | 0.340 | 3.172 | 3.203 | 51.9 | 1.037 | 3.160 | 0.060 | 0.332 |
| | Mumbai | 0.075 | 3.499 | 3.476 | 68.9 | 1.081 | 2.705 | 0.211 | 0.136 |
| | New Delhi | 0.062 | 3.515 | 3.491 | 62.5 | 1.083 | 2.696 | 0.197 | 0.119 |
| | Osaka | 0.243 | 3.298 | 3.306 | 51.0 | 1.025 | 3.155 | 0.069 | 0.292 |
| | Phnom Penh | 0.324 | 3.193 | 3.235 | 81.6 | 1.040 | 2.784 | 0.205 | 0.188 |
| | Pyongyang | 0.024 | 3.557 | 3.568 | 132.4 | 1.097 | 2.524 | 0.294 | 0.120 |
| | Seoul | 0.009 | 3.573 | 3.573 | 53.5 | 1.048 | 3.011 | 0.101 | 0.205 |
| | Shanghai | 0.121 | 3.447 | 3.433 | 233.0 | 1.040 | 3.017 | 0.156 | 0.317 |
| | Singapore | 0.005 | 3.578 | 3.570 | 64.7 | 1.077 | 2.994 | 0.110 | 0.215 |
| | Sydney | 0.092 | 3.480 | 3.431 | 93.1 | 1.073 | 2.674 | 0.206 | 0.087 |

| Region | City | | | | | | | | |
|---|---|---|---|---|---|---|---|---|---|
| | Taipei | 0.158 | 3.402 | 3.428 | 73.5 | 1.068 | 3.096 | 0.110 | 0.305 |
| | Tokyo | 0.050 | 3.528 | 3.529 | 49.6 | 1.046 | 2.950 | 0.119 | 0.186 |
| | Ulaanbaatar | 0.058 | 3.519 | 3.463 | 88.7 | 1.065 | 2.486 | 0.283 | 0.061 |
| Europe | Amsterdam | 0.071 | 3.504 | 3.488 | 65.8 | 1.080 | 2.897 | 0.146 | 0.205 |
| | Athens | 0.041 | 3.538 | 3.532 | 55.5 | 1.019 | 3.245 | 0.056 | 0.363 |
| | Barcelona | 0.108 | 3.462 | 3.460 | 78.1 | 1.052 | 3.135 | 0.078 | 0.303 |
| | Berlin | 0.011 | 3.572 | 3.570 | 113.1 | 1.040 | 3.002 | 0.118 | 0.259 |
| | Budapest | 0.050 | 3.528 | 3.516 | 93.0 | 1.032 | 3.037 | 0.096 | 0.231 |
| | Copenhagen | 0.029 | 3.552 | 3.551 | 78.0 | 1.048 | 2.881 | 0.146 | 0.194 |
| | Dublin | 0.024 | 3.557 | 3.541 | 71.5 | 1.061 | 2.492 | 0.279 | 0.068 |
| | Glasgow | 0.047 | 3.531 | 3.513 | 72.3 | 1.079 | 2.620 | 0.238 | 0.109 |
| | Helsinki | 0.006 | 3.577 | 3.571 | 42.0 | 1.063 | 2.348 | 0.395 | 0.134 |
| | Kiev | 0.014 | 3.568 | 3.554 | 125.1 | 1.053 | 2.813 | 0.164 | 0.160 |
| | Lisbon | 0.023 | 3.558 | 3.546 | 60.8 | 1.068 | 2.923 | 0.108 | 0.154 |
| | London | 0.015 | 3.566 | 3.564 | 70.3 | 1.061 | 2.561 | 0.251 | 0.070 |
| | Madrid | 0.019 | 3.562 | 3.553 | 62.5 | 1.050 | 3.079 | 0.065 | 0.210 |
| | Moscow | 0.007 | 3.576 | 3.573 | 130.5 | 1.055 | 2.999 | 0.074 | 0.170 |
| | Munich | 0.078 | 3.496 | 3.482 | 96.0 | 1.046 | 2.958 | 0.099 | 0.200 |
| | Oslo | 0.008 | 3.574 | 3.564 | 78.0 | 1.095 | 2.711 | 0.197 | 0.113 |
| | Paris | 0.016 | 3.566 | 3.568 | 71.5 | 1.023 | 3.110 | 0.050 | 0.240 |
| | Prague | 0.049 | 3.529 | 3.513 | 84.5 | 1.065 | 2.807 | 0.177 | 0.171 |
| | Reykjavik | 0.056 | 3.522 | 3.529 | 63.2 | 1.071 | 2.540 | 0.283 | 0.117 |
| | Rome | 0.005 | 3.578 | 3.578 | 73.7 | 1.070 | 2.820 | 0.161 | 0.145 |
| | Sarajevo | 0.039 | 3.540 | 3.558 | 94.7 | 1.133 | 2.522 | 0.270 | 0.078 |
| | Stockholm | 0.006 | 3.577 | 3.568 | 82.0 | 1.091 | 2.681 | 0.222 | 0.141 |
| | Venice | 0.017 | 3.564 | 3.553 | 23.2 | 1.090 | 2.474 | 0.300 | 0.073 |
| | Vienna | 0.050 | 3.528 | 3.515 | 90.4 | 1.043 | 2.985 | 0.122 | 0.244 |
| | Warsaw | 0.036 | 3.544 | 3.532 | 90.9 | 1.043 | 2.717 | 0.204 | 0.160 |
| Latin America | Bogota | 0.040 | 3.539 | 3.529 | 58.4 | 1.044 | 2.977 | 0.122 | 0.234 |
| | Buenos Aires | 0.151 | 3.411 | 3.423 | 104.8 | 1.011 | 3.548 | 0.027 | 0.576 |
| | Caracas | 0.029 | 3.551 | 3.564 | 95.3 | 1.148 | 2.710 | 0.217 | 0.145 |
| | Havana | 0.029 | 3.551 | 3.552 | 86.9 | 1.040 | 3.130 | 0.118 | 0.357 |
| | Lima | 0.278 | 3.254 | 3.228 | 76.7 | 1.017 | 3.161 | 0.040 | 0.331 |
| | Mexico City | 0.154 | 3.408 | 3.406 | 69.9 | 1.043 | 2.977 | 0.146 | 0.264 |
| | Port au Prince | 0.028 | 3.552 | 3.554 | 55.0 | 1.088 | 2.495 | 0.295 | 0.087 |
| | Rio de Janeiro | 0.014 | 3.568 | 3.566 | 74.0 | 1.055 | 2.804 | 0.172 | 0.147 |
| | Sao Paulo | 0.002 | 3.581 | 3.580 | 76.0 | 1.050 | 2.936 | 0.120 | 0.176 |
| Middle East/Africa | Baghdad | 0.083 | 3.490 | 3.498 | 68.3 | 1.033 | 3.043 | 0.050 | 0.144 |
| | Beirut | 0.206 | 3.344 | 3.308 | 63.9 | 1.026 | 3.061 | 0.072 | 0.218 |
| | Cairo | 0.041 | 3.538 | 3.526 | 66.6 | 1.067 | 2.996 | 0.085 | 0.171 |
| | Cape Town | 0.025 | 3.556 | 3.553 | 75.2 | 1.102 | 2.793 | 0.183 | 0.162 |
| | Casablanca | 0.094 | 3.477 | 3.461 | 48.0 | 1.048 | 3.026 | 0.080 | 0.178 |
| | Damascus | 0.043 | 3.536 | 3.525 | 65.8 | 1.085 | 2.801 | 0.146 | 0.107 |
| | Dubai | 0.031 | 3.550 | 3.529 | 79.7 | 1.087 | 2.925 | 0.074 | 0.073 |
| | Istanbul | 0.007 | 3.576 | 3.574 | 50.1 | 1.059 | 2.998 | 0.093 | 0.174 |
| | Jerusalem | 0.014 | 3.568 | 3.562 | 44.0 | 1.092 | 2.735 | 0.180 | 0.109 |
| | Johannesburg | 0.019 | 3.562 | 3.556 | 88.6 | 1.098 | 2.865 | 0.158 | 0.182 |
| | Lagos | 0.039 | 3.540 | 3.521 | 87.2 | 1.048 | 2.619 | 0.223 | 0.070 |
| | Mogadishu | 0.375 | 3.123 | 3.292 | 39.4 | 1.019 | 3.346 | 0.055 | 0.472 |
| | Nairobi | 0.014 | 3.568 | 3.556 | 91.8 | 1.083 | 2.506 | 0.279 | 0.075 |
| | Tehran | 0.137 | 3.427 | 3.405 | 52.0 | 1.045 | 2.652 | 0.240 | 0.134 |
| US/Canada | Atlanta | 0.315 | 3.204 | 3.197 | 112.5 | 1.074 | 2.806 | 0.164 | 0.153 |
| | Baltimore | 0.223 | 3.324 | 3.367 | 100.0 | 1.036 | 3.182 | 0.085 | 0.360 |
| | Boston | 0.026 | 3.554 | 3.552 | 77.0 | 1.039 | 2.945 | 0.135 | 0.211 |
| | Charlotte | 0.002 | 3.582 | 3.581 | 117.2 | 1.067 | 2.546 | 0.288 | 0.139 |

| | | | | | | | | |
|---|---|---|---|---|---|---|---|---|
| Chicago | 0.899 | 2.083 | 2.103 | 105.3 | 1.016 | 3.343 | 0.074 | 0.507 |
| Cleveland | 0.486 | 2.961 | 2.899 | 103.7 | 1.029 | 2.979 | 0.091 | 0.198 |
| Dallas | 0.305 | 3.218 | 3.182 | 106.1 | 1.042 | 3.120 | 0.091 | 0.317 |
| Denver | 0.678 | 2.634 | 2.571 | 102.7 | 1.031 | 3.249 | 0.071 | 0.416 |
| Detroit | 0.582 | 2.807 | 2.718 | 101.2 | 1.012 | 3.352 | 0.053 | 0.482 |
| Honolulu | 0.034 | 3.545 | 3.550 | 101.8 | 1.073 | 2.681 | 0.252 | 0.185 |
| Houston | 0.425 | 3.052 | 3.006 | 96.2 | 1.045 | 3.027 | 0.127 | 0.307 |
| Las Vegas | 0.542 | 2.874 | 2.775 | 86.1 | 1.079 | 2.676 | 0.230 | 0.166 |
| Los Angeles | 0.348 | 3.161 | 3.145 | 109.9 | 1.048 | 2.911 | 0.171 | 0.273 |
| Manhattan | 0.669 | 2.650 | 2.571 | 82.2 | 1.017 | 3.508 | 0.027 | 0.572 |
| Miami | 0.811 | 2.341 | 2.291 | 96.7 | 1.023 | 3.236 | 0.069 | 0.407 |
| Minneapolis | 0.749 | 2.486 | 2.464 | 115.4 | 1.023 | 3.393 | 0.053 | 0.521 |
| Montreal | 0.204 | 3.346 | 3.332 | 87.4 | 1.057 | 3.239 | 0.051 | 0.344 |
| New Orleans | 0.123 | 3.444 | 3.457 | 99.6 | 1.035 | 3.378 | 0.077 | 0.526 |
| Orlando | 0.481 | 2.969 | 2.929 | 100.1 | 1.064 | 2.914 | 0.120 | 0.237 |
| Philadelphia | 0.312 | 3.209 | 3.267 | 83.9 | 1.030 | 3.315 | 0.047 | 0.398 |
| Phoenix | 0.586 | 2.801 | 2.563 | 97.1 | 1.073 | 2.795 | 0.186 | 0.171 |
| Pittsburgh | 0.018 | 3.564 | 3.565 | 94.0 | 1.054 | 2.854 | 0.173 | 0.231 |
| Portland | 0.679 | 2.632 | 2.680 | 82.1 | 1.041 | 3.032 | 0.146 | 0.327 |
| San Francisco | 0.278 | 3.253 | 3.226 | 94.4 | 1.033 | 3.304 | 0.087 | 0.454 |
| Seattle | 0.723 | 2.542 | 2.474 | 97.2 | 1.028 | 3.107 | 0.136 | 0.369 |
| St Louis | 0.276 | 3.256 | 3.225 | 107.0 | 1.023 | 3.165 | 0.098 | 0.374 |
| Toronto | 0.474 | 2.980 | 2.885 | 103.1 | 1.090 | 2.994 | 0.109 | 0.217 |
| Vancouver | 0.749 | 2.488 | 2.413 | 103.7 | 1.022 | 3.308 | 0.073 | 0.455 |
| Washington | 0.377 | 3.121 | 3.113 | 99.5 | 1.038 | 3.252 | 0.065 | 0.370 |

NOTE: $\varphi$ is the orientation-order indicator, $H_o$ represents street orientation entropy, $H_w$ represents weighted street orientation entropy, $\bar{l}$ represents median street segment length (meters), $\varsigma$ represents average circuity, $\bar{k}$ represents average node degree, $P_{de}$ represents the proportion of nodes that are dead-ends, and $P_{4w}$ represents the proportion of nodes that are four-way intersections.

Venice, Mogadishu, Helsinki, Jerusalem, and Casablanca have the shortest median street segment lengths (indicating fine-grained networks) while Kiev, Moscow, Pyongyang, Beijing, and Shanghai have the longest (indicating coarse-grained networks). Due to their straight gridded streets, Buenos Aires, Detroit, and Chicago have the least circuitous networks (only 1.1%-1.6% more circuitous than straight-line distances), while Caracas, Hong Kong, and Sarajevo have the most circuitous networks (13.3%-14.8% more circuitous than straight-line distances) due largely to topography. Helsinki and Bangkok have the lowest average node degrees, each with fewer than 2.4 streets per node. Buenos Aires and Manhattan have the greatest average node degrees, both over 3.5 streets per node. Buenos Aires and Manhattan similarly have the largest proportions of four-way intersections and the smallest proportions of dead-end nodes.

Figure 1 and Table 2 aggregate these results by world region (though note that the regional aggregation sample sizes are relatively small and thus the usual caveats apply). On average, the US/Canadian cities exhibit the lowest street orientation entropy, circuity, and proportions of dead-ends as well as the highest median street segment lengths, average node degrees, and proportions of four-way intersections. They are also by far the most grid-like in terms of $\varphi$. On average, the European cities exhibit the highest street orientation entropy and proportion of dead-ends as well as the lowest average node degrees. They are the least grid-like in terms of $\varphi$. To illustrate the geography of these order/entropy trends, Figure 2 maps the 100 study sites by $\varphi$ terciles. As expected, most of the sites in the US and Canada fall in the highest tercile (i.e., they have low entropy and highly-ordered, grid-like street orientations), but the notable exceptions of high-entropy Charlotte, Boston, and Pittsburgh fall in the lowest tercile. Most of the sites in Europe fall in the lowest tercile (i.e., they have high entropy and disordered street orientations). Most of the sites across the Middle East and South Asia fall in the middle tercile.

**Table 2.** Mean values of indicators aggregated by world region.

|  | $\varphi$ | $H_o$ | $H_w$ | $\tilde{I}$ | $\varsigma$ | $\bar{k}$ | $P_{de}$ | $P_{4w}$ |
|---|---|---|---|---|---|---|---|---|
| Asia/Oceania | 0.123 | 3.439 | 3.437 | 80.6 | 1.062 | 2.836 | 0.171 | 0.184 |
| Europe | 0.033 | 3.547 | 3.540 | 78.7 | 1.061 | 2.814 | 0.172 | 0.172 |
| Latin America | 0.081 | 3.490 | 3.489 | 77.5 | 1.055 | 2.971 | 0.140 | 0.257 |
| Middle East/Africa | 0.081 | 3.490 | 3.490 | 65.8 | 1.064 | 2.883 | 0.137 | 0.162 |
| US/Canada | 0.427 | 3.003 | 2.969 | 98.8 | 1.043 | 3.090 | 0.116 | 0.334 |

NOTE: See Table 1 for column definitions.

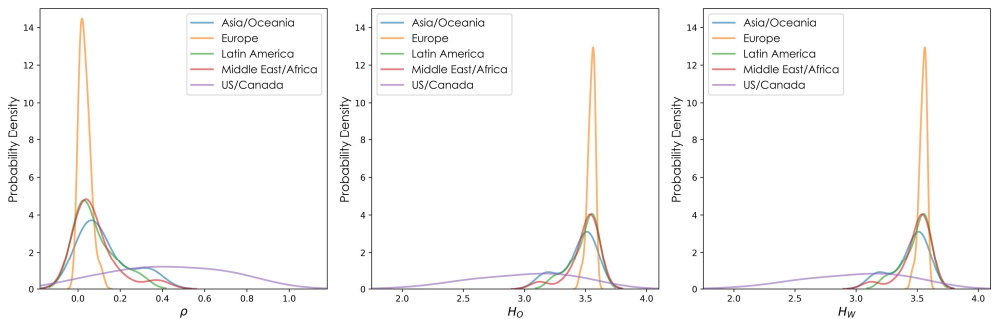

**Figure 1.** Probability densities of cities' $\varphi$, $H_o$, and $H_w$, by region, estimated with kernel density estimation. The area under each curve equals 1.

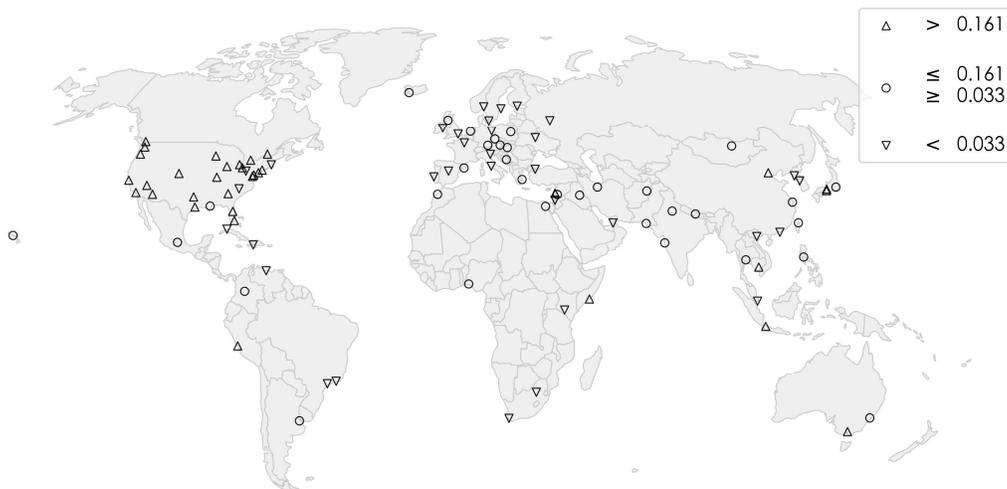

**Figure 2.** Map of study sites in terciles of orientation-order, $\varphi$.

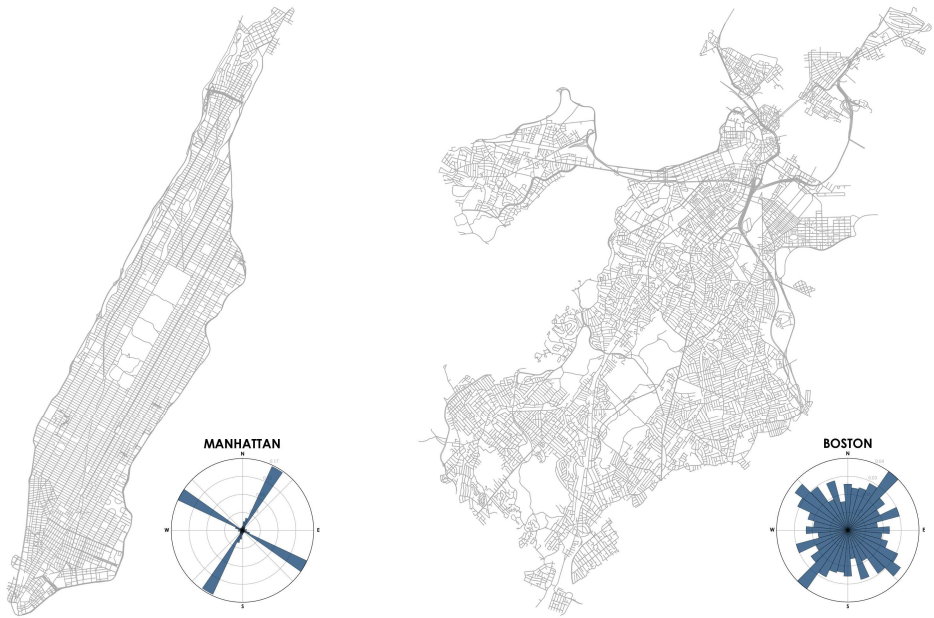

**Figure 3.** Street networks and corresponding polar histograms for Manhattan and Boston.

To better visualize spatial order and entropy, we plot polar histograms of each city's street orientations. Each polar histogram contains 36 bins, matching the description in the methods section. Each histogram bar's direction represents the compass bearings of the streets (in that histogram bin) and its length represents the relative frequency of streets with those bearings. The two examples in Figure 3 demonstrate this. On the left, Manhattan's 29° angled grid originates from the New York Commissioners' Plan of 1811, which laid out its iconic 800-foot × 200-foot blocks (Ballon, 2012; Koeppel, 2015). Broadway weaves diagonally across it, revealing the path dependence of the old Wickquasgeck Trail's vestiges, by which Native Americans traversed the island long before the first Dutch colonists arrived (Holloway, 2013). On the right, Boston features a grid in some neighborhoods like the Back Bay and South Boston, but they tend to not align with one another, resulting in the polar histogram's jumble of competing orientations. Furthermore, the grids are not ubiquitous and Boston's other streets wind in various directions, resulting from its age (old by American standards), terrain (relatively hilly), and historical annexation of various independent towns with their own pre-existing street networks.

Figures 4 and 5 visualize each city's street orientations as a polar histogram. Figure 4 presents them alphabetically to correspond with Table 1 while Figure 5 presents them in descending order of $\varphi$ values to better illustrate the connection between entropy, griddedness, and statistical dispersion. The plots exhibit perfect 180° rotational symmetry and, typically, approximate 90° rotational symmetry as well. About half of these cities (49%) have an at least approximate north-south-east-west orientation trend (i.e., 0°-90°-180°-270° are their most common four street bearing bins). Another 14% have the adjacent orientations (i.e., 10°-100°-190°-280° or 80°-170°-260°-350°) as their most common. Thus, even cities without a strong grid orientation often still demonstrate an overall tendency favoring north-south-east-west orientation (e.g., as seen in Berlin, Hanoi, Istanbul, and Jerusalem).

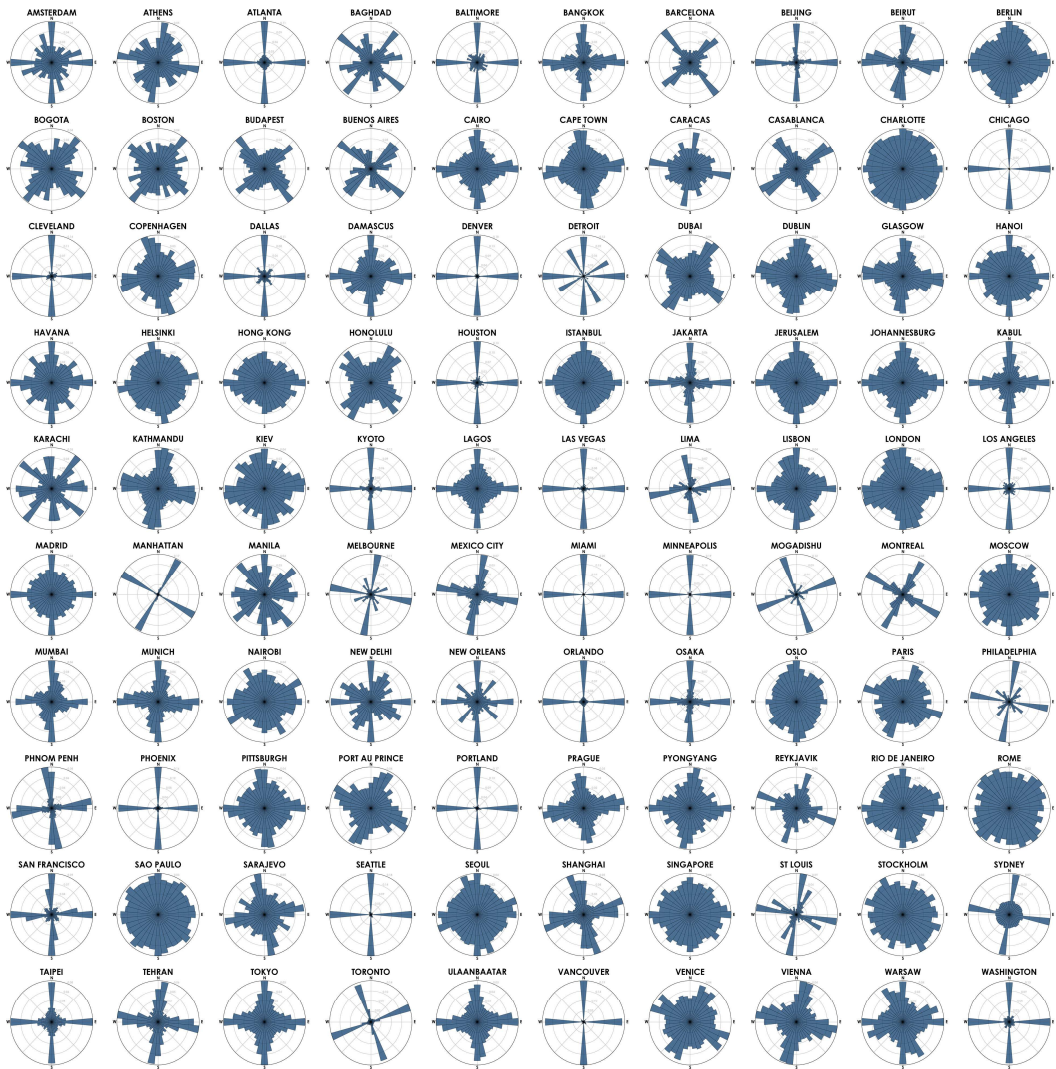

**Figure 4.** Polar histograms of 100 world cities' street orientations, sorted alphabetically corresponding with Table 1.

Straightforward orthogonal grids can be seen in the histograms of Chicago, Miami, and others. Detroit presents an interesting case, as it primarily comprises two separate orthogonal grids, one a slight rotation of the other. While Seattle's histogram looks fairly grid-like, it is not fully so: most of Seattle is indeed on a north-south-east-west grid, but its downtown rotates by both 32° and 49° (Speidel, 1967). Accordingly, there are observations in all of its bins and its $H_o = 2.54$ and $\varphi = 0.72$, whereas a perfect grid would have $H_o = 1.39$ and $\varphi = 1$. Thus, it is about 72% of the way between perfect disorder and a single perfect grid. However, its rotated downtown comprises a relatively small number of streets such that the rest of the city's much larger volume swamps the histogram's relative frequencies. The same effects are true of similar cites, such as Denver and Minneapolis, that have downtown grids at an offset from the rest of the city (Goodstein, 1994). If an entire city is on a grid except for one relatively small district, the primary grid tends to overwhelm the fewer offset streets (cf. Detroit, with its two distinct and more evenly-sized separate grids).

Figures 4 and 5 put Chicago's low entropy and Charlotte's high entropy in perspective. Of these 100 cities, Chicago exhibits the closest approximation of a single perfect grid with the majority of its streets falling into just four bins centered on 0°, 90°, 180°, and 270°. Its $\varphi = 0.90$, suggesting it is 90% of the way between perfect disorder and a single perfect grid, somewhat remarkable for such a large city. Most American cities' polar histograms similarly tend to cluster in at least a rough, approximate way. Charlotte, Rome, and São Paulo, meanwhile, have nearly uniform distributions of street orientations around the compass. Rather than one or two primary orthogonal grids organizing city circulation, their streets run more evenly in every direction.

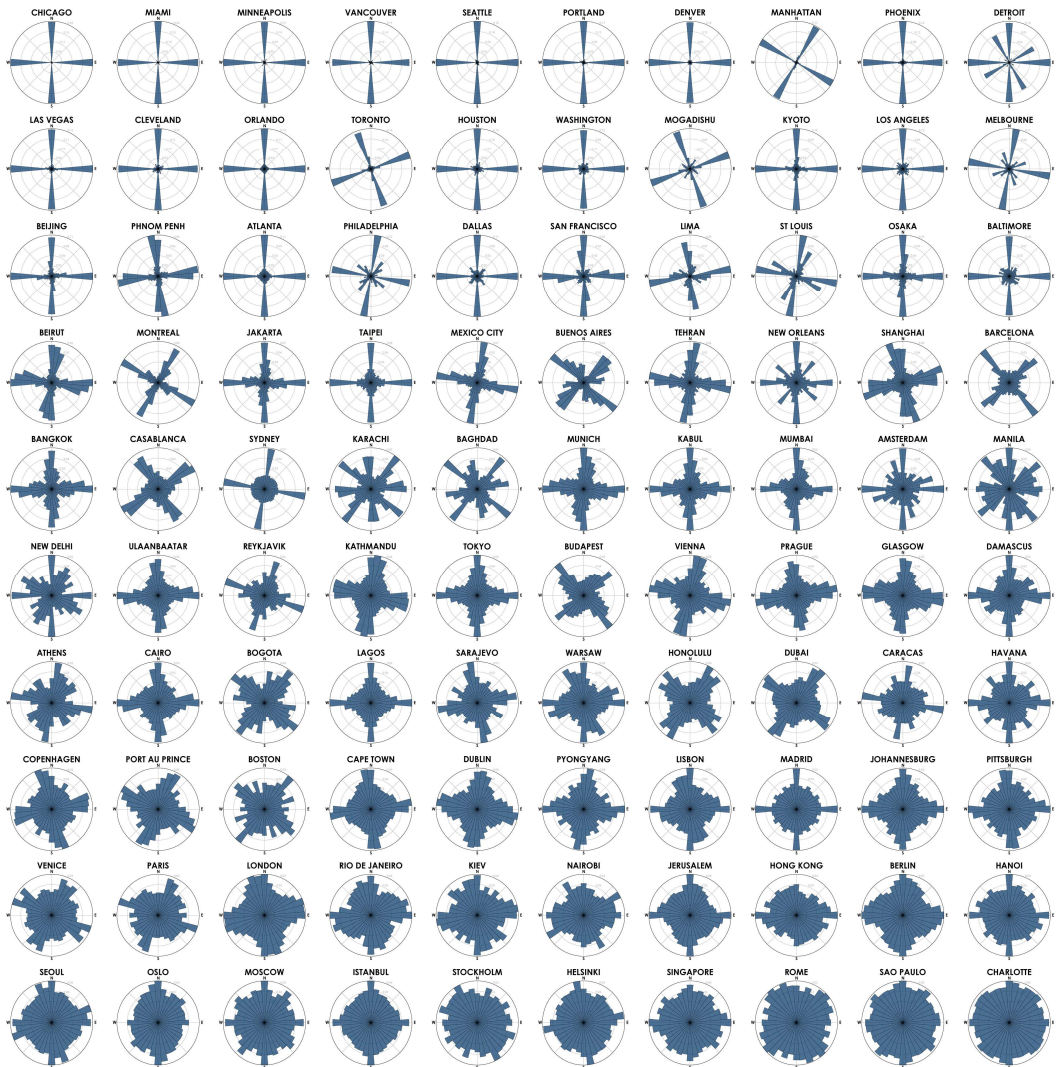

**Figure 5.** Polar histograms from Figure 4, resorted by descending $\varphi$ from most to least grid-like (equivalent to least to greatest entropy).

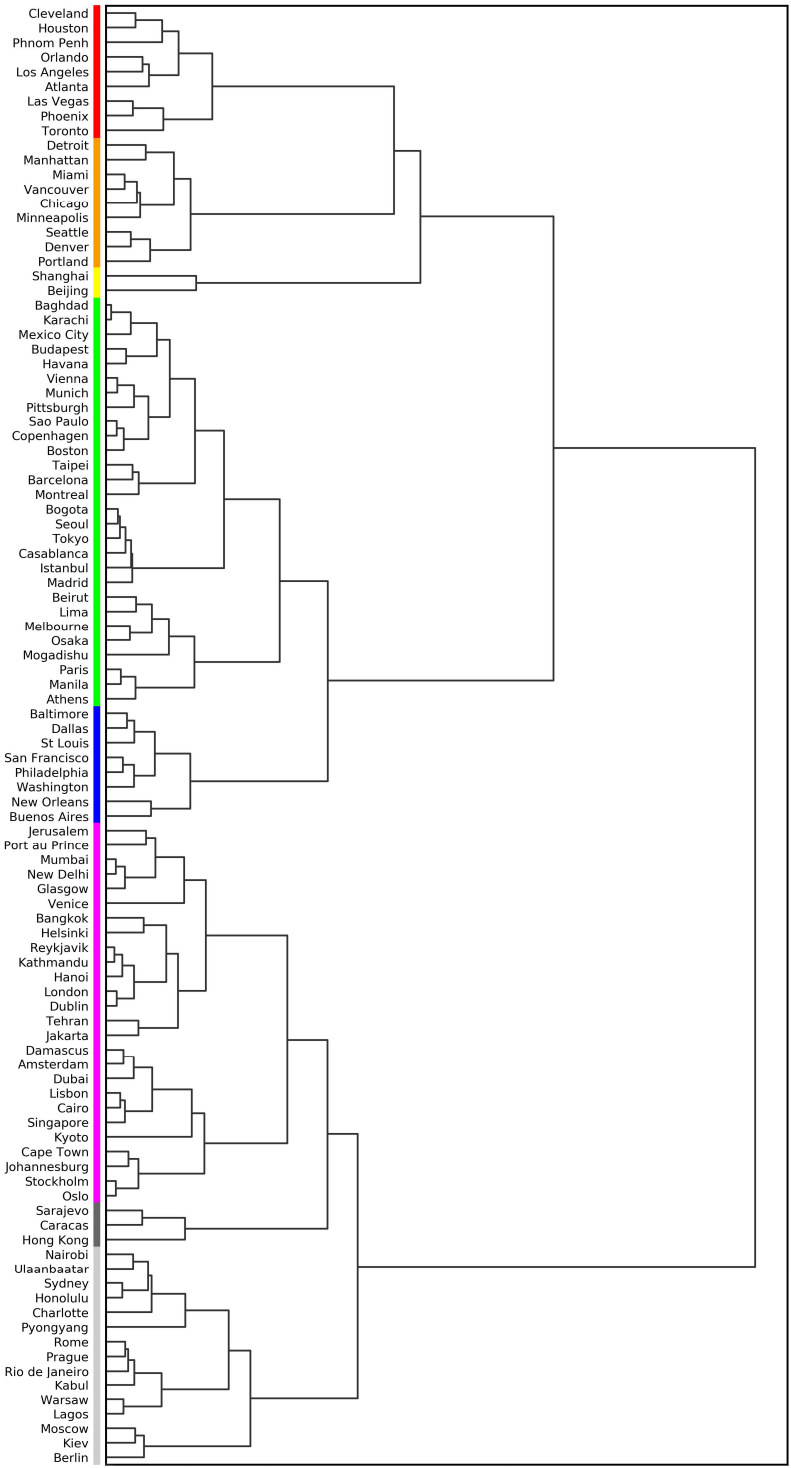

**Figure 6.** Cluster analysis dendrogram. Cluster colors correspond to Figure 7.

As discussed earlier, orientation entropy and weighted orientation entropy are strongly correlated. Additionally, $\varphi$ moderately and negatively correlates with average circuity ($r(\varphi, \varsigma) = -0.432$, $p < 0.001$) and the proportion of dead-ends ($r(\varphi, P_{de}) = -0.376$, $p < 0.001$), and moderately and positively correlates with the average node degree ($r(\varphi, \bar{k}) = 0.518$, $p < 0.001$) and proportion of four-way intersections ($r(\varphi, P_{4w}) = 0.634$, $p < 0.001$). As hypothesized, cities with more grid-like street orientations tend to also have more streets per node, more four-way junctions, fewer winding street patterns, and fewer dead-ends. Besides these relationships, $\varphi$ also has a weak but significant correlation with median street segment length ($r(\varphi, \tilde{l}) = 0.27$, $p < 0.01$), concurring with previous findings examining the UK alone (Gudmundsson and Mohajeri, 2013). Average circuity moderately strongly and negatively correlates with the average node degree ($r(\varsigma, \bar{k}) = -0.672$, $p < 0.001$) and the proportion of four-way intersections ($r(\varsigma, P_{4w}) = -0.689$, $p < 0.001$). Cities with more winding street patterns tend to have fewer streets per node and fewer grid-like four-way junctions.

Figure 6 presents the dendrogram obtained from the cluster analysis, allowing us to systematically explore cities that are more- or less-similar to each other. The dendrogram's structure suggests three high-level superclusters of cities, but for further analysis, we cut its tree at an intermediate level (eight clusters) for better external validity and more nuanced insight into those larger structures. To visualize these clusters another way, we map their four-dimensional feature space to two dimensions using t-SNE, a manifold learning approach for nonlinear dimensionality reduction that is well-suited for embedding higher-dimensional data in a plane for visualization (Maaten and Hinton, 2008). Figure 7 scatterplots the cities in these two dimensions: the t-SNE projection preserves their cluster structure relatively well despite inherent information loss, but, given the global density-equalizing nature of the algorithm, the relative distances within and between clusters are not preserved in the embedding and should not be interpreted otherwise.

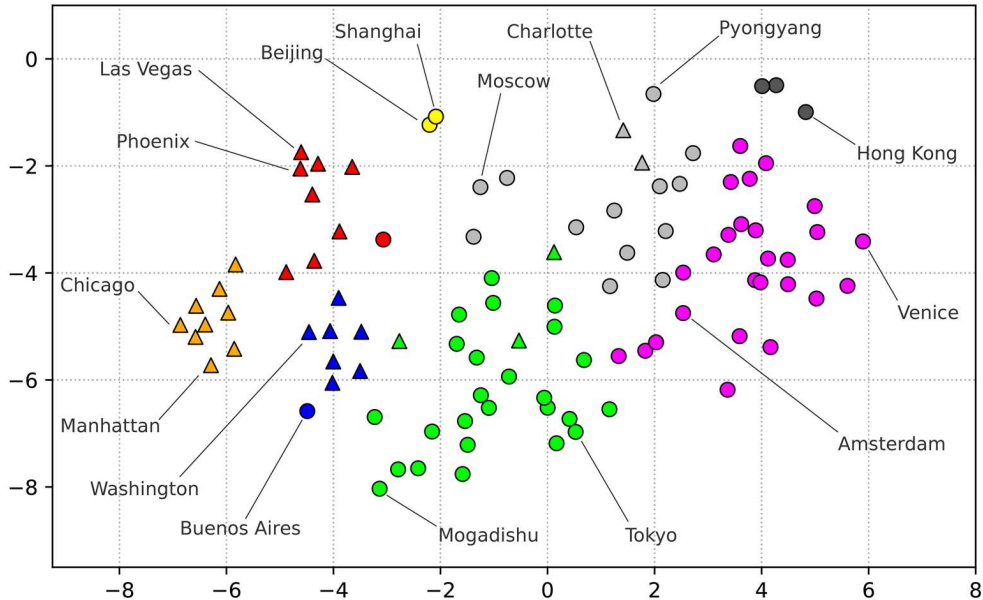

**Figure 7.** Scatterplot of cities in two dimensions via t-SNE. Cluster colors correspond to Figure 6. Triangles represent US/Canadian cities and circles represent other cities.

Most of the North American cities lie near each other in three adjacent clusters (red, orange, and blue), which contain grid-like—and almost exclusively North American—cities. The orange cluster represents relatively dense, gridded cities like Chicago, Portland, Vancouver, and Manhattan. The blue cluster contains less-perfectly gridded US cities, typified by San Francisco and Washington (plus, interestingly, Buenos Aires). The red cluster represents sprawling but relatively low-entropy cities like Los Angeles, Phoenix, and Las Vegas. Sprawling, high-entropy Charlotte is in a separate cluster (alongside Honolulu) dominated by cities that developed at least in part under the auspices of 20th century communism, including Moscow, Kiev, Warsaw, Prague, Berlin, Kabul, Pyongyang, and Ulaanbaatar. Beijing and Shanghai are alone in their own cluster, more dissimilar from the other study sites. The dark gray cluster comprises the three cities with the most circuitous networks: Caracas, Hong Kong, and Sarajevo. While the US cities tend to group together in the red, orange, and blue clusters, the other world regions' cities tend to distribute more evenly across the green, purple, and light gray clusters.

## Discussion

The urban design historian Spiro Kostof once said: "We 'read' form correctly only to the extent that we are familiar with the precise cultural conditions that generated it… The more we know about cultures, about the structure of society in various periods of history in different parts of the world, the better we are able to read their built environment" (Kostof, 1991, p. 10). This study does not identify whether or how a city is planned or not. Specific spatial logics cannot be conflated with planning itself, which takes diverse forms and embodies innumerable patterns and complex structures, as do informal settlements and organic urban fabrics. In many cities, centrally planned and self-organized spatial patterns coexist, as the urban form evolves over time or as a city expands to accrete new heterogeneous urban forms through synoecism.

Yet these findings do, in concert, illustrate different urban spatial ordering principles and help explain some nuances of griddedness. For example, gridded Buenos Aires has a $\varphi$ value suggesting it only follows a single grid to a 15% extent. However, its low circuity and high average node degree values demonstrate how it actually comprises multiple competing grids—which can indeed be seen in Figures 4 and 5—and it clusters accordingly in Figures 6 and 7 with gridded American cities. Jointly considered, the $\varphi$ indicator, average circuity, average node degree, and median street segment length tell us about the extent of griddedness and its character (curvilinear, straight-line, monolithic, heterogeneous, coarse-grained, etc.). Charlotte further illustrates the importance of taking these indicators together. Although its $\varphi$ and orientation entropy are more similar to European cities' than American cities', it is of course an oversimplification to claim that Charlotte is therefore the US city with the most "European" street network—in fact, its median street segment length is about 50% longer than that of the average European city, and among European cities, Charlotte clusters primarily with those of the Communist Bloc. Pittsburgh, on the other hand, sits alone in a small sub-cluster with Munich and Vienna.

We find that cities with higher $\varphi$ values also tend to have higher node degrees, more four-way intersections, fewer dead-ends, and less-winding street patterns. That is, cities that are more consistently organized according to a grid tend to exhibit greater connectedness and less circuity. Interestingly, the $H_o$ and $H_w$ orientation entropies are extremely similar and strongly correlated: the weighted curvatures (versus straight-line orientation) of individual street segments have little impact on citywide orientation entropy, but the average circuity of the city network as a whole positively correlates with orientation entropy. This finding deserves further exploration.

These results also demonstrate substantial regional differences around the world. Across these study sites, US/Canadian cities have an average $\varphi$ value nearly thirteen-times greater than that of European cities, alongside nearly double the average proportion of four-way intersections. Meanwhile, these European cities' streets on average are 42% more circuitous than those of the US/Canadian cities. These findings illustrate the differences between North American and European urban patterns. However, likely due to such regional heterogeneity, this study finds statistical relationships somewhat weaker (though still significant) than prior findings examining cities in the UK exclusively.

Accordingly, given the heterogeneity of these world regions, future research can estimate separate statistical models for individual regions or countries—or even the neighborhoods of a single city to draw these findings closer to the scale of planning/design practice. The methods and indicators developed here offer planners and designers a toolbox to quantify urban form patterns and compare their own cities to those elsewhere in the world. Our preliminary results suggest trends and patterns, but future work should introduce additional controls to clarify relationships and make these findings more actionable for researchers and practitioners. For instance, topography likely constrains griddedness and influences circuity and orientation entropy: a study of urban elevation change and hilliness in conjunction with entropy and circuity would help clarify these relationships. Additionally, further research can unpack the relationship between development era, design paradigm, city size, transportation planning objectives, and street network entropy to explore how network growth and evolution affect spatial order. Finally, given the importance of taking multiple indicators in concert, future work can develop a grid-index to unify them and eventually include streetscape and width attributes as further enrichment to explore walkability and travel behavior.

## Conclusion

Street networks organize and constrain a city's transportation dynamics according to a certain spatial logic—be it planned or unplanned, ordered or disordered. Past studies of this spatial order have been challenged by small samples, limited geographies, and abstract entropy indicators. This study accordingly looked at a larger sample of cities around the world, empirically examining street network configuration and entropy across 100 cities for the first time. It measured network orientation entropy, circuity, connectedness, and grain. It also developed an orientation-order indicator $\varphi$, to quantify the extent to which a network is ordered according to a single grid.

This study found significant correlations between $\varphi$ and other indicators of spatial order, including street circuity and measures of connectedness. It empirically confirmed that the cities in the US and Canada are more grid-like (exhibiting far less entropy and circuity) than was typical elsewhere. It is noteworthy that Chicago—the foremost theoretical model of 20[th] century city growth and development in urban studies (Dear, 2001; Park and Burgess, 1925; Wirth, 1928)—is an extreme outlier among world cities in terms of spatial orientation-order. In sum, these methods and indicators demonstrate scalable techniques to empirically measure and visualize the complexity of spatial order, illustrating patterns in urbanization and transportation around the world.